\documentclass{article}
\usepackage[utf8]{inputenc}
\usepackage{xcolor}
\usepackage{soul}
\usepackage{url}
\usepackage{comment}
\usepackage{graphics}
\usepackage{graphicx}
\usepackage{listings}
\usepackage{authblk}
\usepackage{hyperref}

\newcommand{\eduard}[1]{{\color{black}{#1}}}

\newcommand{\newmaterial}[1]{{\color{black}{#1}}}

\newcommand{\roberto}[1]{{\color{magenta}{#1}}}
\usepackage{xurl}

\title{Serverless Computing: A Security Perspective}

\author[1]{Eduard Marin}
\author[1]{Diego Perino}
\author[2]{Roberto Di Pietro}
\affil[1]{Telef\'{o}nica Research, Spain}
\affil[2]{Hamad Bin Khalifa University

College of Science and Engineering, Doha-Qatar}

\date{}

\begin{document}

\maketitle

\section*{Abstract}

Serverless Computing is a virtualisation-related paradigm that promises to simplify application management and to solve the last challenges in the field: scale down and easy to use. The implied cost reduction, coupled with a simplified management of underlying applications, are expected to further push the adoption of virtualisation-based solutions, including cloud-computing or telco-cloud solutions. However, in this quest for efficiency, security is not ranked among the top priorities, also because of the (misleading) belief that current solutions developed for virtualised environments could be applied (as is) to this new paradigm. Unfortunately, this is not the case, due to the highlighted idiosyncratic features of serverless computing. In this paper, we review the current serverless architectures, abstract and categorise their founding principles, and provide an in depth analyse of them from the point of view of security, referring to principles and practices of the cybersecurity domain. In particular, we show the security shortcomings of the analysed serverless architectural paradigms, point to possible countermeasures, and highlight a few research directions.
\\\\*
{\bf keywords:} Serverless Computing, Architectures, Security\\\\


\section{Introduction}


Virtualisation technologies have played a crucial role for the wide adoption and success of cloud computing. They allowed cloud providers to simultaneously share their resources with many users by placing their applications (i.e., monoliths) inside virtual machines (VMs), offering strong isolation guarantees while providing users \newmaterial{an (apparently)} infinite amount of resources, readily available when their applications need them. The cited features, together with a pay-per-use business model that has contributed to lower the \eduard{total cost of ownership} (TCO), have made cloud computing the most successful computing paradigm of the last decade. However, \eduard{this success also} came with some drawbacks: the need for the users to directly \newmaterial{manage the VMs being one of them}.

New programming models \eduard{have recently} emerged that drastically change the way software developers develop and manage applications for the cloud. \eduard{One such programming} model relies on decomposing an application into multiple, autonomous, limited scope and loosely coupled components\eduard{---also known as \emph{microservices}---}that can communicate with each other via standard APIs. Unfortunately, due to their long startup time and high resource usage, VMs were proven to be inefficient for running microservices. To tackle the cited limitation\eduard{s}, several container technologies (e.g., Docker) were proposed as a lighter alternative. Containers offer increased portability, lower start up time, and \eduard{provide} greater resource utilisation compared to VMs, simplifying the \eduard{development and} management of large-scale applications in the cloud. \eduard{These advantages have led} cloud providers to adopt container technologies and \eduard{to} use them in combination with orchestration platforms, such as Kubernetes \eduard{or Docker Swarm}, \eduard{to fully automate} the deployment, scaling, and management of microservice-based applications. 
\eduard{Nevertheless, similarly to when VMs are used,} \eduard{the microservices paradigm} \eduard{also} requires \eduard{users} to configure and manage the \eduard{underlying} containers (e.g., \eduard{related} libraries and software dependencies). \eduard{In addition,} \eduard{the microservices paradigm} relies on a static billing model where users pay a fixed amount according to the allocated resources and not the resources actually consumed. \eduard{This renders} microservices either unsuitable or not viable for certain types of applications.

{\em Serveless paradigm.}
To tackle the above limitations, a novel \eduard{computing} paradigm has been conceived. \emph{Function-as-a-Service (FaaS)}~\cite{serverless1, serverless2} allows software developers to outsource all infrastructure management and operational tasks to cloud providers, making it possible for them to focus solely on writing the code for their applications~\cite{DBLP:journals/corr/abs-1902-03383}. FaaS is also widely known as serverless computing---in the following we will mainly use this term. Serverless is the most widely known realisation of FaaS to date (although FaaS could in principle be developed in other ways too). It is worth noting that the term ``serverless'' does not mean that there are no servers, but rather that software developers do not need to worry about configuring and managing them.

{\em Serverless advantages.} With serverless, the application logic is divided into a set of \emph{small} and \emph{stateless} functions, each running within a separate execution environment (e.g., a container) and performing a single task. Functions are typically \emph{short-lived} and are invoked relatively infrequently via \eduard{events}, while storage is provided by separate cloud services shared across users. \eduard{This way,} serverless decouples storage from computation, so that these \eduard{two elements} can be provisioned, managed, and priced separately. \newmaterial{Furthermore}, the cloud provider is now the one responsible for automatically and transparently spawning and managing function instances in worker nodes as well as performing all operational tasks. These latter tasks include server and OS maintenance, patching, logging, load balancing or auto-scaling. Unlike prior cloud \eduard{paradigms}, serverless adopts a pure pay-per-use model where users are only billed based on the resources (e.g., CPU, network or memory) they consume, significantly reducing application deployment cost. In addition to the advantages that serverless \eduard{offers} to software developers in terms of flexibility, scalability, performance and costs, it also \eduard{brings} important benefits to cloud providers. As functions are invoked only occasionally and are executed for a short period of time, cloud providers can achieve a high\eduard{er} degree of co-location in their servers and a more optimal use of their resources. These two latter points, when carefully planned and orchestrated, can result in an even more profitable model for cloud providers. 

{\em Serverless players.} Cloud providers, such as Amazon~\cite{lambda}, Microsoft~\cite{microsoftazure}, Google~\cite{googlefunctions}, \eduard{or Alibaba~\cite{alibaba}}, are already offering serverless computing services to their customers. Meanwhile, the research community has also developed several open-source serverless platforms such as OpenFaaS~\cite{OpenFaaS}, Knative~\cite{knative}, or Kubeless~\cite{kubeless}.

{\em Serverless security issues.} With the increase in volume and diversity of attacks against the cloud, security and privacy will be a key factor that, if not addressed, could hamper the widespread adoption of serverless computing. At first glance, one could argue that serverless computing is intrinsically more secure than its predecessors \eduard{because of} its characteristics (e.g., the short duration of functions), or due to the fact that it could inherit security features already developed for other virtualisation solutions.
However, as we will show in the next sections, serverless brings many new, unique security challenges \eduard{that open the door for new types of security attacks}\footnote{\newmaterial{For an overview of application-level attacks against serverless, we refer the reader to a recent report by OWASP~\cite{OWASP}.}}. \eduard{It is also worth noting that} implementing serverless applications requires a major change in mindset from software developers, \eduard{not only} in the way applications are written \eduard{but also} in the way they are protected from security attacks. As of today, serverless security is still a relatively new field, with only a few initial works specifically addressing the security issues of the serverless ecosystem.


{\bf Contribution.}
In this paper, we provide several contributions. We first review and categorise state of the art serverless solutions; later, we analyse pros and cons of the introduced architectural categories; further, we assess, from a security perspective, the fundamental principles of the main revised architectural choices. Finally, starting from the highlighted weaknesses, we provide several research directions, appealing to Industry, Academia, and practitioners, to further enhance the security of the serverless ecosystem as a whole.

\section{Background and Related Work}


In this section we revise the current serverless ecosystem, analysing the five main elements any serverless platform is composed of, and then discussing the currently available security solutions.

\subsection{Serverless ecosystem}
\begin{figure}
  \centering
  \includegraphics[width=0.85\linewidth]{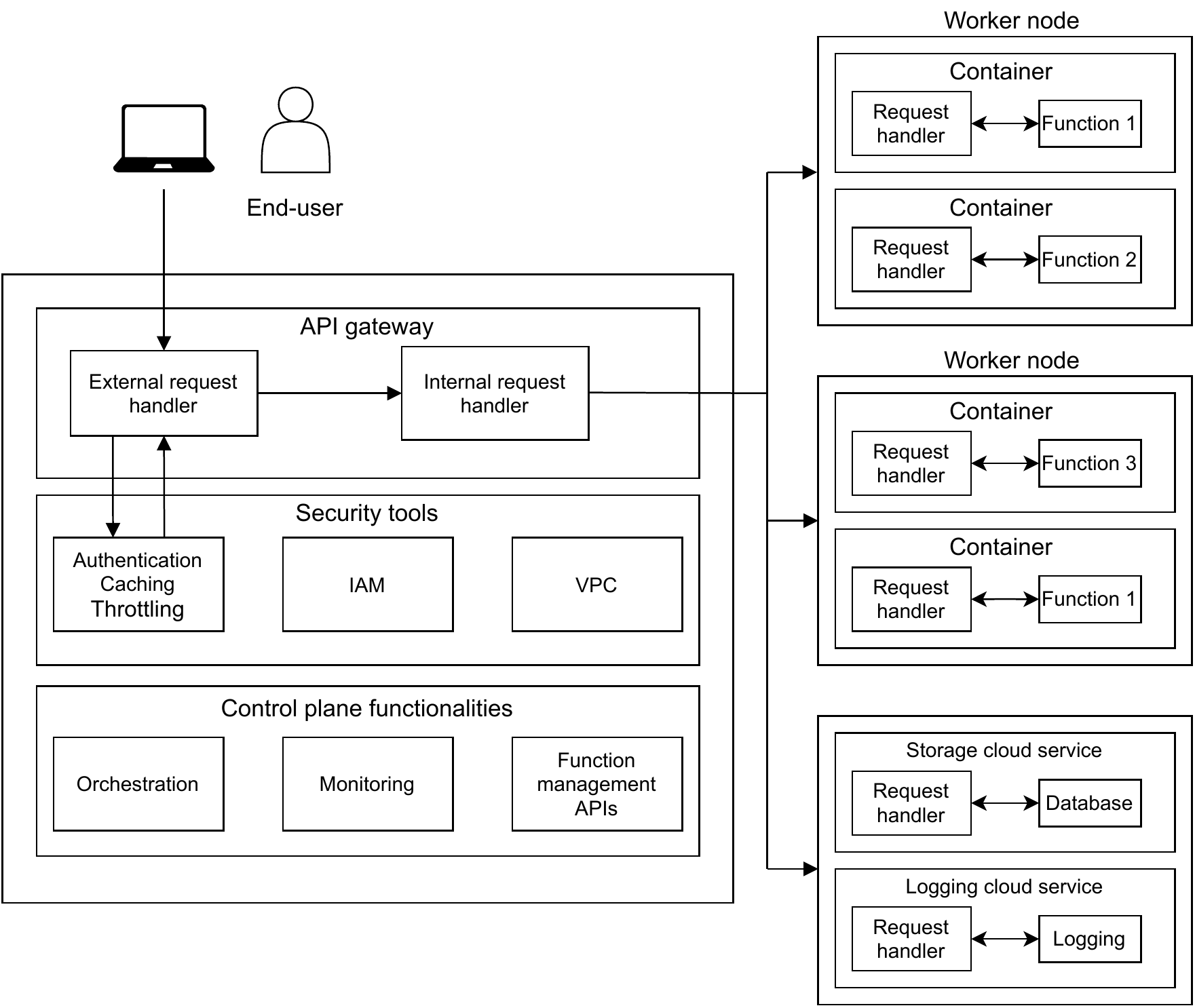}
  \caption{Serverless platform} 
  \label{fig1}
\end{figure}

As shown in Figure~\ref{fig1}, a serverless platform \eduard{is comprised of} (at least) the following five elements: (i) functions; (ii) API gateways; (iii) (shared) cloud services;\\ (iv) security tools; and, (v) control plane. 

{\em Functions.} Functions are the \eduard{core} component of serverless platforms. \newmaterial{They can be written in many different programming languages~\cite{proglangu}}. Software developers can either write the\eduard{m} themselves, rely on open-source third-party functions (e.g.,~\cite{repo}), or use proprietary functions for which they \eduard{must} pay licensing fees. 
Functions are typically run inside a newly-generated, isolated execution environment within a worker node. \eduard{They} are executed in response to distinct external and internal events specified by application owners (e.g., HTTP requests, modification to objects in storage, table updates, or function transitions). \eduard{It is worth noting that not all the defined functions can communicate with the outside world}. 

{\em API gateways.} API gateways provide a central management service that exposes REST API endpoints and act as a bridge between end-users and functions. All communications between end-users and functions \eduard{always} have to first traverse API gateways. API gateways support various security mechanisms for throttling, caching, authenticating and authorising incoming API calls. The latter can be done in \eduard{a variety of ways}, e.g., \newmaterial{relying on external} identity providers or \newmaterial{specifying a range of} IP addresses from which requests \newmaterial{can originate}. 

{\em Shared cloud services.}
Serverless platforms integrate a wide range of cloud services available to application owners used to extend the \eduard{functions' capabilities}, e.g., to collect various types of data (e.g., \eduard{using} Kinesis) or to achieve long- and short-term storage (e.g., \eduard{using} S3 or DynamoDB).

{\em Security tools for software developers.} Cloud providers also make a set of tools to ease \newmaterial{workflows} security management \eduard{available to software developers}. Prominent examples are Identity and Access Management (IAM) and Virtual Private Cloud (VPC). The former \eduard{allows the configuration of} fine-grained access controls to \newmaterial{authenticate and restrict the resources functions have access to}, whereas the latter enables the creation of private, isolated networks for secure communications between applications \newmaterial{that belong to the same organisation}. 

{\em Control plane functionalities.} Serverless platforms comprise \eduard{multiple} control plane functionalities for cloud providers to operate, \eduard{manage}, and monitor their infrastructures. For example, \eduard{there is} an orchestrator component \eduard{that} handles the process of assigning functions to worker nodes. \eduard{Another example is the} monitoring component used to periodically check the status of worker nodes and their execution environments. This way, if a failure is detected, the affected functions can be quickly instantiated in other worker nodes.

To sum up, serverless platforms are complex and dynamic ecosystems with many distinct components. To design a secure serverless ecosystem, one must consider the security provided by each of its components and their interplay.

\subsection{Existing infrastructure-level security controls}
Current serverless platforms typically rely on a combination of proprietary and open-source (built into the Linux kernel) security mechanisms to protect functions and the underlying infrastructure against attacks. \newmaterial{These security mechanisms can be clustered into the following five categories}:


{\em Host hardening.} The execution environments currently used by cloud providers are middle-ground solutions between traditional VMs and containers, where functions have access to some parts of the host OS kernel. \eduard{Unfortunately, \eduard{by exploiting a single vulnerability in the OS kernel}, adversaries can} trigger memory leaks, compromise \eduard{co-resident} functions or, even worse, gain elevated privileges in the host.
\eduard{To mitigate such attacks,} several Linux kernel features and modules, such as \emph{SELinux}, \emph{apparmor}, \emph{seccomp} and \emph{seccomp-bpf}, are used to
define and enforce a security policy that specifies the allowed behaviour for a function in terms of system calls, arguments, and host resources accessed.

{\em Isolation between functions.} To achieve isolation among functions running on the same \eduard{worker node}, each function is run inside a separate execution environment with its own dedicated \emph{namespace}. Through the use of namespaces, all processes running within an execution environment have their own filesystem, network, processes, inter-process communication, and devices. This \eduard{separation mechanism} prevents processes residing in an execution environment from listing or tampering with processes in other execution environments, making it harder for adversaries to launch attacks. Moreover, \emph{cgroups} are used to control the amount of resources (e.g., CPU, memory or disk I/O) each execution environment can use. This \eduard{mechanism} ensures that each execution environment gets a fair share of the available resources, thwarting attacks where adversaries aim to over-utilise resources to leave other execution environments without sufficient resources.

{\em Data at rest (Storage).} Every time software developers upload a new function to a serverless platform, the cloud provider encrypts the function's code using standard cryptographic algorithms (e.g., AES-GCM~\cite{AES}). This is done to protect the function's code from unauthorised access while at rest. Similarly, any user data stored or handled by cloud services is kept encrypted while at rest as well. The cryptographic keys \eduard{employed} are stored and managed securely using specific cloud services (e.g., AWS Key Management Service) \eduard{which often rely on} hardware security modules (or similar technologies) to \eduard{provide a higher level of security}.

{\em Network.} Whenever the cloud provider instantiates a function in an execution environment of a worker node, the function's code is securely transported to it using standard cryptographic protocols (e.g., TLS).
\eduard{All function-to-function and function-to-service communications are protected in transit also using secure standard cryptographic protocols like TLS}. Moreover, security mechanisms \eduard{are put} in place such that it is impossible for external users to initiate inbound/outbound network communications with a function without passing by the API gateway first. To protect functions from internal threats, application owners can \eduard{create their own} virtual, isolated networks \eduard{(using iptables and routing tables)}, \eduard{allowing them to} control \eduard{the way} their functions communicate with each other and with the outside world.

{\em Access control.} Cloud providers \eduard{typically} offer \eduard{multiple} mechanisms to authenticate and authorise external API calls before passing the requests to the corresponding functions. While \eduard{these methods} alleviate some attacks by external adversaries, \eduard{they do not} prevent abuses \eduard{by} malicious functions. \eduard{For the latter}, cloud providers rely on IAM to let application owners restrict functions' privileges as much as possible.

\section{Threat model}

We mainly distinguish between two types of adversaries: (i) \emph{external}; and, (ii) \emph{internal}. External adversaries (i.e., malicious end-users) typically carry out their attacks from outside the cloud by leveraging \newmaterial{user-controlled input fields in} any of the existing \newmaterial{event} APIs serverless platforms \eduard{offer}. These attacks can enable adversaries to \newmaterial{run any arbitrary commands inside the function in order to retrieve sensitive data (e.g., session tokens stored in environment tables) or} tamper with the execution of any function or cloud service that receives \newmaterial{maliciously-crafted} input data and does not apply proper sanitation techniques. \newmaterial{While some injection attacks, such as those that exploit cross-site scripting or the ones based on code/command injection, are well-known (as they are applicable to standard web applications), serverless functions can also be triggered from many different event sources, which broadens their attack surface significantly~\cite{eventinjection2, eventinjection}}.\\*

On the contrary, internal adversaries refer to \eduard{adversaries} who \eduard{have full control of one (or more) functions and} conduct attacks from inside the cloud. \newmaterial{In the case of public clouds, it is relatively easy for adversaries to deploy functions in serverless platforms in order to attempt to perform attacks from inside.} \eduard{Such adversaries} can attempt to: \eduard{(i) create cover channels~\cite{10.1145/3442381.3450100,10.5555/3277355.3277369};} (ii) \newmaterial{conduct privilege escalation attacks (e.g., to compromise other co-resident functions or worker nodes)}~\cite{priviesca}; (iii) retrieve or tamper with sensitive data (e.g., data in storage services); (iv) gather knowledge about runtime and infrastructure\newmaterial{~\cite{10.5555/3277355.3277369}}; or, (v) conduct various types of Denial-of-Service (DoS) attacks\newmaterial{~\cite{10.5555/3277355.3277369}}. 

\newmaterial{Though out of the scope of this paper, it is worth mentioning that from a privacy point of view there is an increasing concern that cloud providers can inadvertently or deliberately reveal sensitive data (e.g., through malicious insiders) to third-parties}. \eduard{Due to this latter threat, within the research community it is common to model} cloud providers as \emph{honest-but-curious} entities, \eduard{meaning that they \eduard{will} run their customers} applications as intended but, at the same time, they \eduard{will} try to learn as much information as possible about the ongoing computations.



In \eduard{the} next sections we analyse the impact of serverless computing on security, discussing the pros and cons of the paradigm in relationship with its contribution to the security posture of the supported ecosystem.

\section{Serverless as a Security Enabler}

In this section, we discuss some principles and use cases related to the inherent advantages of serverless from a security point of view.

\paragraph{\eduard{Increased difficulty in carrying out attacks} \newmaterial{and reduced attack impact.}} The fact that serverless functions are stateless and short-lived significantly raises the bar for adversaries to successfully execute their attacks. \newmaterial{Indeed,} serverless imposes strict limits on the time available to adversaries \newmaterial{for them} to retrieve sensitive data from functions or to move laterally \eduard{in order} to perform more sophisticated attacks. \newmaterial{This is important because experience has shown that adversaries who compromise servers often remain undetected for very long periods, carrying out malicious activities at a very slow pace, not to generate signals that could lead to detection -- this type of attack being labelled advanced persistent threat (APT). The consequences of such long-lasting attacks can be severe, ranging from intellectual property theft (e.g., trade secrets or patents), compromised sensitive information (e.g., employee and user private data) to total site takeovers. With serverless, long-standing servers do not exist, thus adversaries must carry out their attacks -- including the reconnaissance phase -- again and again, increasing both the attack costs and the risks of being detected.} \newmaterial{Additionally, by using small single-purpose functions to realise the applications, serverless allows not only the definition of more fine-grained security policies, but also a significant reduction of the impact of attacks, as now adversaries who compromise a function can only exploit the capabilities of such function. Finally, due to the way applications are designed within the serverless paradigm, not all serverless functions need to be reachable from the Internet, hampering adversaries from conducting some attacks (e.g., those that aim to exfiltrate sensitive data)}. 

\paragraph{Less security responsibilities for software developers.} In contrast to prior cloud programming models where software developers play an important role in guaranteeing the security of their applications, \newmaterial{serverless security is a shared responsibility between software developers and the cloud provider.} In serverless, one can distinguish between \emph{``security of the cloud''} and \emph{``security in the cloud''}. ``Security of the cloud'' is the responsibility of cloud providers and encompasses all measures in place to keep the underlying infrastructure and cloud services (e.g., the execution environments on which functions run or the virtualisation layer) secure from adversaries. \newmaterial{Although software developers have less control and require trust in the chosen cloud provider, delegating all infrastructure-related security tasks to cloud providers is considered to be an effective mechanism to eliminate a wide number of attacks. Providing, maintaining and operating an infrastructure that is secure by design is the core business model of cloud providers offering serverless and hence it is one of their main focuses.}

\newmaterial{Instead, ``security in the cloud'' is the responsibility of software developers and refers to the security mechanisms employed to: (1) prevent vulnerabilities in the functions; (2) protect the application's data (stored in cloud services); and, (3) secure the entire workflows (e.g., guaranteeing that all functions are executed only with the minimum privileges needed). This can be achieved by leveraging cloud services and tools that cloud providers make available to software developers which allow the control and management of access to resources, monitoring components, logging information, verifying network configurations, protecting against DDoS attacks, implementing firewalls, inspecting traffic, or securing access control and key management (among others).} The cited concepts are critical ones, and need to be fully seized by software developers, the alternative being the developers ignoring the consideration of security for their applications, or to make unrealistic assumptions about the security measures put in place by cloud providers---in both cases, a dreadful scenario.







\paragraph{Resistance to traditional Denial-of-Service attacks.} \newmaterial{Serverless, by construction, enjoys elasticity (as it can adapt to workload changes by provisioning and de-provisioning resources)}, 
\newmaterial{thanks to its efficient and automatic auto-scaling}. As such, serverless platforms \newmaterial{provide increased resistance against} \newmaterial{many different types of DoS/DDoS attacks whose goal is to overwhelm network bandwidth, trigger many compute-heavy actions in parallel, or exploit flaws in the application in order to cause infinite loops.} \newmaterial{While auto-scaling has already been used in previous computing paradigms (e.g., microservices), before serverless this feature required the usage of an external service which was complex to use that software developers had to configure manually. In serverless, auto-scaling is considerably simpler, more effective and less costly, making it a very important feature in serverless-based applications.} 



\section{Serverless as a Security Risk} 

In this section, we detail several aspects of serverless that can negatively affect security. \eduard{Table~\ref{table:securitycomparison} compares the level of security offered by the serverless paradigm against the one provided by its predecessors for a few interesting dimensions.}



\begin{table}[ht]
	\caption{Security comparison between monolith applications, microservices, and \\ serverless.} \small
	\centering
	\resizebox{\columnwidth}{!}{%
	\begin{tabular}{|c|c|c|c|}
		\hline
		 & \textbf{Monoliths} & \textbf{Microservices} & \textbf{Serverless} \\ [0.5ex]
		\hline 
		Feasibility of long-lasting attacks & High & High & Low \\ \hline 
		Main responsible for security & Mostly app owners & Mostly app owners & Shared responsibility \\ \hline 
		Entry points for adversaries & Few & \newmaterial{Medium} & Many \\ \hline 
		Resistance to Denial-of-Service attacks & Low & Medium & High \\ \hline 
		Denial-of-Wallet attacks & Not possible & Not possible & Possible \\ \hline 
		Communication with other components & None & Medium & High \\ \hline 
		Visibility of underlying infrastructure & High & High & Low \\ \hline
		\hline 
	\end{tabular} 
	}
 \label{table:securitycomparison}
	\end{table}

\paragraph{1) Larger attack surface} 
Serverless computing \eduard{exposes} a significantly larger attack surface compared to its predecessors \eduard{for three main reasons}. First, as functions are stateless and are only intended to perform a single task, they are required to constantly interact with other functions and (shared) cloud services. However, the definition and enforcement of security policies specifying which functions and cloud services can be accessed by each function in such dynamic and complex environments is very challenging\newmaterial{~\cite{10.1145/3366423.3380173, 10.1145/3427228.3427665}}. Second, functions can be triggered by many external and internal event sources (e.g., 47 \newmaterial{event types} in Amazon Lambda~\cite{functionevents} \newmaterial{and 11 event types in Azure~\cite{azureserverless}}) with multiple formats and encoding. \newmaterial{This creates many possible entry-points for adversaries to gain control of functions; even more than when using microservices due to the fact that serverless applications are stateless and event-driven}. Third, serverless platforms include a number of new components and cloud services, many of which are shared across users. \newmaterial{Again, the fact that serverless functions are stateless, simple and event-driven, together with the fact that cloud providers want to provide greater application performance and achieve a much more optimal use of their resources, means that serverless platforms include many more components that are shared between users with respect to previous computing paradigms.} Such shared components \eduard{can enable} new forms of side or covert channels that \eduard{can allow adversaries to} retrieve sensitive data or \eduard{violate the specified security policy}.
\newmaterial{Next, we present three types of attacks that could be launched against serverless platforms which deserve to be investigated in depth by the research community.}

\begin{itemize} 
\item \newmaterial{\textbf{Side channel attacks.} Adversaries could exploit the way serverless platforms are designed and implemented in order to conduct new forms of side channel attacks. For example, they could leverage weaknesses in the execution environments where functions are executed in order to obtain host-system state information (e.g., power consumption or performance data) or individual process execution information, such as process scheduling, cgroups or process running status. All this information could help adversaries to uniquely identify a worker node or a function instance as well as to perform more effective and efficient attacks. Similarly, the sequence of functions traversed in response to external events triggered by users could also reveal information to adversaries (e.g., the role of the person triggering the request). As functions are triggered in response to an action performed by a user, adversaries could gain insights about the users by looking at the functions' metadata (e.g., when or how often functions are called). Other more sophisticated side channels could be devised, based on the fact that there exist many components and cloud services shared across users. Adversaries are particularly interested in any shared component subject to a change in its state based on the processed data---since these components could leak sensitive data about users and functions through a side channel. It is important to note that side channel attacks in the context of serverless computing have not yet been fully investigated by the scientific community. Therefore, an in-depth evaluation is needed to analyse the feasibility, extent and consequences of such attacks and to propose effective countermeasures to defend against them.}

\item \newmaterial{\textbf{Race conditions.} Serverless platforms can be vulnerable to attacks caused by inconsistencies in any component whose functionality is distributed across several nodes or that contains multiple replicas. For example, let us assume that software developers decide to modify the code of a function while several replicas of this function are running. In such a case, there could be a (small) time window where the serverless platform is in an inconsistency state: some incoming requests would be handled by the old version of the function while some others could get served by the new version of the function. Such inconsistencies could be caused, for example, by cloud providers reusing execution environments with the old version of the function for a certain period of time. Adversaries could abuse such undesirable behaviour to conduct security attacks with the goal of accessing or modifying data that otherwise might no longer be available to them. Similar attacks could also be carried out when other parameters are modified (e.g., IAM roles, memory sizes or environment variables) while multiple replicas of the same function are executed. Modifying these parameters at runtime can lead to race conditions that adversaries can exploit to lower the overall security of the serverless platform. While race conditions can also happen in a microservices architecture, the smaller granularity offered by serverless platforms increases the risk of inconsistencies across function versions. Overall, we believe this research area deserves more attention from the scientific community, both to understand the security threats and to design effective countermeasures against them.}


\item \newmaterial{\textbf{Billing attacks.} Though serverless offers increased protection against traditional DoS/DDoS attacks, these attacks can lead to a new, serverless-specific attack that takes advantage of the fact that application owners are billed based on the amount of resources their functions consume. By sending many requests to functions, adversaries can now perform so called Denial-of-Wallet (DoW) attacks \newmaterial{\cite{DoW}} with the purpose of significantly increasing the cost for application owners. Although some mitigating countermeasures already exist against DoW attacks (e.g., creating a billing alert to notify application owners if they exceed a predefined spending limit), these attacks are not easy to defend against and require additional control measures: first, to detect abnormal behaviour; and, later to discriminate which legitimate invocations to allow, and which ones to drop. The uniqueness of serverless in this context, is the fact that invocation and billing are at a very small granularity, i.e., the function; an adversary can simply generate invocations to a function in order to perform the attack while other auto-scaling constructions would require the generation of a high load on a full container or VM. As such, the consequences of successfully launching such attacks can be more severe when targeting serverless platforms. Moreover, given the fact that computation can evolve only via function calls, blocking legitimate function invocations would represent a more serious threat than that experienced by the cited auto-scaling twins of serverless. }


\end{itemize}


\paragraph{Proprietary cloud provider infrastructures} Cloud providers are now the ones responsible for conducting all operational and infrastructure tasks, including those aimed to protect their infrastructures and the hosted applications from internal and external threats. Unfortunately, \eduard{they} typically keep most information about their infrastructures confidential, making it difficult for security experts to scrutinise the security and privacy of serverless platforms. Within the security \newmaterial{research} community, \newmaterial{it is widely known that relying on} security-through-obscurity \newmaterial{alone} \newmaterial{is} a dangerous approach \eduard{that} may conceal insecure designs~\cite{Ross}. 
\eduard{Motivated by the above rationales,} researchers have devoted significant efforts into reverse-engineering and documenting how the serverless platforms of the main cloud providers were developed in an attempt to understand their core design decisions \newmaterial{(e.g.,~\cite{10.5555/3277355.3277369, 9284261})}. 
\eduard{Yet, there are still many components within serverless platforms that remain unexplored to date and hence whose security level is unknown}.


\paragraph{Security vs.\ performance vs.\ cost} Ideally, cloud providers would like to develop serverless platforms that jointly maximise the security and performance of their infrastructures while \eduard{maximising their revenue and} keeping the incurred cost as low as possible. \newmaterial{However, experience has shown that cloud providers, when it comes to which dimension to curb in order to keep cost under control, do not have security at the top of their priority list of features to preserve.}


\begin{table}[ht]
	\caption{Comparison of execution environments features. Note that by ``containers" and ``VMs" we mean cold containers and traditional VMs, respectively. g-Visor refers to Google's sandboxed containers while microVMs corresponds to Amazon's VMs created by Firecracker.} \small
	\centering
	\resizebox{1.05\columnwidth}{!}{%
	\begin{tabular}{|c|c|c|c|c|}
		\hline
		 & \textbf{VMs} & \textbf{Containers} & \textbf{g-Visor} & \textbf{microVMs}\\ [0.5ex]
		\hline 
 	 \newmaterial{Number of functionalities in host OS kernel~\cite{10.1145/3381052.3381315}} & Almost none & Almost all & Less than in containers & More than in VMs \\ \hline 
		 App startup times & Very high & Medium & Medium & High \\ \hline 
		 Isolation guarantees & \newmaterial{Medium-high} & Low & Medium & \newmaterial{High} \\ \hline 
		 Complexity & High & Medium-low & Medium-low & \newmaterial{Medium} \\ \hline 
		 Written in safe prog. languages & No & No & Yes (Golang) & Yes (Rust)\\ \hline 

		\hline 
	\end{tabular} 
	}
 \label{table:nonlin}
	\end{table}

\begin{itemize}

 \item \textbf{Cold containers vs.\ warm containers.} When serverless was first introduced, it was proposed to create a new, fresh container (or similar technology) isolated from others \eduard{whenever} a function is invoked. However, repeatedly booting a function from scratch inside a newly-generated container (i.e., a {\em cold} container) can be an expensive operation latency-wise. Most \eduard{serverless} functions are executed only for a very short period of time \eduard{and hence} the container's booting latency \eduard{would be} similar to the function's execution time. This \eduard{would be problematic} for applications with stringent latency requirements. Another reason why the use of cold containers is an issue \newmaterial{(from the point of view of the cloud provider)} is that customers are not billed for the time it takes for their containers to boot. To overcome the limitations posed by cold containers, cloud providers have opted for using so-called {\em warm} containers, i.e., containers that are reused to run multiple instances of the same function. Warm containers not only reduce the functions' startup times but also improve efficiency, e.g., by keeping and reusing local caches or maintaining long-lived connections between invocations. However, the advantages warm containers offer come at the \newmaterial{expense} of providing \eduard{fewer} security guarantees. While warm containers restore the default values in the filesystem within the execution environment every time the function is completed, they contain a small writable {\em /tmp/} disk space to share state across different function invocations. Adversaries who compromise a function, could leverage the fact that the data in /tmp/ is kept across all invocations to execute long-lasting attacks without raising any alarm\newmaterial{~\cite{60mili,ser,tmpfolder}}. To avoid such attacks, application owners can opt for disabling the possibility of reusing the same execution environment to run the same function multiple times. This is particularly useful for those execution environments running functions that perform security-sensitive tasks. Yet, disabling warm containers may not always be a viable option since this can introduce a significant overhead in terms of application performance---\newmaterial{which introduces a penalty in contexts, like Telcos, where stringent response times have to be assured. In case the usage of warm containers is required to meet the application performance requirements, one possible way to mitigate this issue would be for cloud providers to reduce the size of the /tmp/ folder to the minimum extent possible and to analyse its contents after every function invocation. The challenge being how to distinguish between the data stored in /tmp/ that comes from the application and malicious code that adversaries could store within this directory.}

 
 
 


 \item \textbf{Execution environments.} The selection of the execution environment \eduard{on which} functions \eduard{are executed} is \eduard{crucial} for cloud providers since \eduard{it} strongly \eduard{impacts} the security and performance \eduard{of} their serverless platform. \eduard{For example, containers entail less overhead and provide greater resource utilisation than VMs\footnote{Note that some VMs programming languages, such as the Java Virtual Machine, are also known to pose a number of security issues.} but this also results in weaker isolation guarantees}. \eduard{Similarly}, it has \eduard{also} been proposed to combine VMs and containers together (i.e., by placing all containers of a user in a VM) to benefit from the isolation guarantees of VMs while maintaining (to some extent) the performance advantages offered by containers. However, \newmaterial{in practice} cloud providers have opted for developing their own execution environments and open-sourcing their code. 
For example, Amazon designed Firecracker~\cite{Firecracker}, a solution that builds upon the KVM hypervisor to create and manage \eduard{so called} microVMs \eduard{through a new virtual machine monitor along with a new API}. Typically, each MicroVM hosts several containers belonging to the same user. Firecracker---used in Lambda since 2018---has been designed with simplicity and minimalism as its key goals, and attempts to reuse existing components wherever possible. With Firecracker, Amazon can not only span new VMs quickly but also run thousands of VMs on each \eduard{worker node} \eduard{securely} with minimal overhead. \eduard{Similarly}, Google has developed \eduard{g-Visor~\cite{gVisor}}, \newmaterial{a user-space application kernel that sits between the containerised application and the host OS and hence provides an additional layer of isolation per container. g-Visor---used in Google's production environment since 2018---implements a substantial portion of the Linux system call interface, allowing g-Visor to intercept and respond to container-invoked system calls. Besides, the interactions between g-Visor and the host OS are restricted, meaning that the host OS attack surface is significantly reduced. Overall, although Firecracker and g-Visor approaches show promise, neither their attack surface nor their security mechanisms have yet been properly evaluated by security experts. Hence, research should focus on understanding their weaknesses and limitations.} 
 
 \textbf{Deterministic vs.\ random scheduling algorithms.} Consider the process adopted by cloud providers to assign functions to worker nodes. The cited task can be fulfilled using deterministic or randomised scheduling algorithms. From a security point of view, randomised scheduling algorithms are preferred over deterministic ones because they offer stronger protection against attacks that could exploit co-residency. 
 However, randomised scheduling algorithms do not consider functional aspects such as worker nodes' resource utilisation or function-to-function communication overhead when choosing the worker nodes that will execute the 
 functions. 
 This leads to a non-optimal allocation of functions that can negatively affect the overall performance of the applications and the underlying serverless infrastructure. To prevent the latter issue, cloud providers typically opt for deterministic scheduling algorithms that lead to a more optimal use of the available resources and less communication overhead. Nevertheless, this approach can be vulnerable to attacks by adversaries that can obtain information about (or tamper with) the way the scheduling algorithms work.

\end{itemize}

\section{Conclusions}


In this contribution we have shown that, on the one hand, serverless computing provides additional security features while, on the other hand, it also introduces unique security threats and challenges---clearly differentiating itself from current virtualisation technologies. 
In particular, we have reviewed current serverless architectures, categorised the current security threats, \eduard{shown actionable hints to improve the current security position}, and highlighted security research directions \eduard{to make serverless the paradigm of choice when looking for virtualisation solutions where security is at a premium.}
 We believe that our contribution, other than being valuable on its own, also paves the way for further research in this domain, a challenging and relevant one for both industry and academia.

\section*{Acknowledgements}
\eduard{We would like to thank the anonymous reviewers for their comments, that helped us to improve the quality of the manuscript.}

\bibliographystyle{acm}
\bibliography{CACM}

\end{document}